\documentclass[prb,twocolumn,showpacs,floatfix]{revtex4}
\usepackage{graphicx}
\usepackage{dcolumn}
\usepackage{bm}

\begin{document}
\title{Wave packet dynamics in 2DEG with spin orbit coupling: splitting and {\bf{\it zitterbewegung}}
}

\author{V.~Ya.~Demikhovskii, G.~M.~Maksimova and E.~V.~Frolova}
\email{demi@phys.unn.ru}
\affiliation{Nizhny Novgorod State
University,\\ Gagarin Ave., 23, Nizhny Novgorod 603950, Russian
Federation}

\date{today}

\begin{abstract}
We study the effect of splitting and {\it zitterbewegung} of 1D
and 2D electron wave packets in the semiconductor quantum well
under the influence of the Rashba spin orbit coupling. Results of
our investigations show that the spin orbit interaction induces
dramatic qualitative changes in the evolution of spin polarized
wave packet. The initial wave packet splits into two parts with
different spin polarization propagating with unequal group
velocity. This splitting appears due to the presence of two
branches of electron spectrum corresponding to the stationary
states with different chirality. It is demonstrated also that in
the presence of external magnetic field ${\bf B}$ perpendicular to
the electron gas plane the wave packet splits into two parts which
rotates with different cyclotron frequencies. It was shown that
after some periods the electron density distributes around
cyclotron orbit and the motion acquire an irregular character. Our
calculations were made for both cases of weak and strong spin
orbit coupling.
\end{abstract}

\pacs{73.21.Hb, 71.10.Pm, 72.10.-d, 73.23.-b}

\maketitle

\section{Introduction}

Producing and detecting spin polarized currents in semiconductor
nonmagnetic devices is the ultimate goal of spintronics. The
intrinsic spin orbit interaction\cite{Shl} existing in low
dimensional systems which couples electron momentum to its spin is
one of the most promising tools for realizing spin polarized
transport. For this reasons, during the last years a substantial
amount of work has been devoted to study effects of spin orbit
interaction on the transport properties of nanostructures (for a
review, see, e. g.\cite{HAE,Zut,ByR}).

At first time the electron wave packet dynamics including the
problem of {\it zitterbewegung} in semiconductor quantum well
under the influence of the spin orbit Rashba and Dresselhaus
coupling has been considered by Schliemann, Loss and Westervelt
\cite{SK}. In this work the oscillatory motion of the electron
wave packets reminiscent of the {\it zitterbewegung} of
relativistic electrons was studied for free electron motion i.e.,
in the absence of electric or magnetic field.  The authors
of\cite{SK} predicted the resonance amplification of {\it
zitterbewegung} oscillations for the electron moving in a quantum
wire with parabolic confinement potential and propose to observe
this fundamental phenomena experimentally using high resolution
scanning probe microscopy imaging techniques.

The {\it zitterbewegung} of the heavy and light holes in 3D
semiconductors was investigated in\cite{SKF}.In this paper the
semiclassical motion of holes in the presence of a constant
electric field was studied by numerical solution of the Heisenberg
equations for momentum and spin operators in the Lattinger model
of spectrum. It was shown that the hole semiclassical trajectories
contain rapid small amplitude oscillations reminiscent the {\it
zitterbewegung} of relativistic electrons. It should be noted,
however, that the spatial structure of the wave packet and the
changing of its shape due to effect of splitting in\cite{SK,SKF}
was not considered.

At the same time the splitting of spin polarized electron beams in
the systems with spin orbit coupling was investigated in a series
of works.  In particular, the authors of papers\cite{Chen,Gon}
propose to use the lateral interface between two regions in gated
two-dimensional heterostructure with different strength of spin
orbit coupling to polarize the electron. They have shown
theoretically that in this structure a beam with a nonzero angle
of incidence splits into some spin polarization components
propagating at different angles. The similar effect of electron
spin-polarized reflection in heterostructures and spatial
separation of the electron beams after reflection has been
observed experimentally in\cite{Rok}.

The transverse electron focusing in systems with spin orbit
coupling at the presence of perpendicular magnetic field was
theoretically analyzed in\cite{Los} where it was shown that in the
weak magnetic field regime and for a given energy, the two
branches of states have different cyclotron radii. The effect of
spatial separation of the electron trajectories of different spin
states in a perpendicular magnetic field has been experimentally
observed in\cite{ZF}.

In this work we study the striking dynamics of the electron wave
packets in a narrow $A_3B_5$  quantum well at the presence of the
spin orbit $k$-linear Rashba coupling, which arise due to
structural inversion ("up-down") asymmetry. The splitting of the
wave packets in two parts appear due to the presence of the
electron states with "plus" and "minus" chirality, which propagate
with different group velocity. These two parts of the split packet
can be characterized by different spin density. It is found that
electron trajectories contain small amplitude damped oscillation.
We show that the packet splitting leads to the damping of {\it
zitterbewegung}.  The splitting and {\it zitterbewegung} of wave
packet is naturally accompanied by its broadening due to effect of
dispersion.

We investigate also the atypical cyclotron dynamics of the wave
packet in a perpendicular magnetic field. It was shown that due to
the spin orbit coupling the packet with spin parallel to the
magnetic field splits into two parts which rotate with different
cyclotron frequencies. We determine the moments when two parts of
the packet are located at opposite points of the cyclotron orbit
and after that they return many times back to their initial state.
With the time due to the incommensurability of the cyclotron
frequencies and the ordinary packet broadening the electron
density distributes randomly around the cyclotron orbit.  All our
calculations were made for the material parameters of the real
semiconductor structures with a relatively strong and weak spin
orbit and Zeeman interaction.

The paper is organized as follows. In Sec. II we introduce the
Green functions for two dimensional electrons in the presence of
Rashba spin orbit interaction and analyze the evolution of 1D wave
packet. The analytical and numerical results illustrate the
effects of packet splitting and {\it zitterbewegung}. In section
III we describe in details the time development of the 2D wave
packets. Finally, in Sec. IV we discus the manifestation of the
spin orbit interaction in the evolution of coherent wave packet in
a magnetic field perpendicular to electron gas plane. The
splitting of the initial coherent packet and distribution electron
probability via cyclotron orbit is considered. Section V concludes
with a discussion of the results. The Appendix provides the
mathematical details necessary to obtain Eqs. (36a) and (36b).

\section{The dynamic of the one-dimensional wave packets}
In this section we consider the specific character of the wave
packet dynamics in the systems with Rashba  spin orbit
coupling\cite{Shl}. The Hamiltonian of the system under
consideration reads $$ H=H_0+H_R=\frac{{\bf p}^2}{2m}+\alpha(\hat
p_y\hat \sigma_x-\hat p_x\hat \sigma_y),\eqno(1) $$ where ${\bf
p}=-i\hbar\nabla$ is the momentum operator, $m$ is the electron
effective mass, $\alpha$ is the Rashba coupling constant, and the
components of the vector ${\bf \sigma}$ denotes the spin Pauli
matrices. The eigenfunctions for the in-plane motion identified by
the quantum numbers ${\bf p}(p_x,p_y)$ are $$ \phi_{{\bf
p},s}({\bf r})=\frac{1}{2\sqrt{2}\pi\hbar}{\rm e}^{i{\bf pr}}
\pmatrix{1\cr -is{\rm e}^{i\varphi}},\eqno(2) $$ Here $\varphi$ is
the angle between the electron momentum $\bf p$ and $x$ axis, so
${\rm e}^{i\varphi}=\frac{p_x+ip_y}{p}$, $s=\pm 1$ denotes the
branch index. The energy spectrum of the Hamiltonian (1)
corresponding to two branches has the form
$$\varepsilon_{\pm}(p)=\frac{p^2}{2m}\pm\alpha p,\eqno(3)$$ where
$p=\sqrt{p_x^2+p_y^2}$. Using the definition ${\bf \hat
v}=\frac{d{\bf r}}{d t}=\frac{i}{\hbar}[H,{\bf r}]$, one can
obtain from Eq.(1) the velocity operator components $$ \hat
v_x=\frac{p_x}{m}-\alpha\sigma_y,\;\hat
v_y=\frac{p_y}{m}+\alpha\sigma_x.\eqno(4) $$

To analyze the time evolution of electron the initial states we
use the Green's function of the nonstationary equation, which is a
non diagonal $2\times2$ matrix $$G_{ik}=\pmatrix{G_{11} &
G_{12}\cr G_{21} & G_{22}}.\eqno(5)$$ Here $i,k=1,2$ are matrix
indexes and matrix elements can be written as an integrals
$$G_{ik}({\bf r},{\bf r }^\prime,t)=\sum\limits_s\int d{\bf
p}\phi_{{\bf p}s,i}({\bf r},t)\phi_{{\bf p}s,k}^\ast({\bf
r}^\prime,0).\eqno(6)$$

In the present section we examine in details the dynamics of the
quasi-1D wave packet in 2D system with spin orbit coupling. This
problem allows the analytical solution. Let at the initial time
$t=0$ wave function to be a plane wave with wave number $p_{0x}$
modulated  by a Gaussian profile and spin polarized along the $z$
direction $$\displaylines{\Psi({\bf
r},0)=\Psi(x,0)=C\exp(-\frac{x^2}{2d^2}+ip_{0x}x/\hbar)\pmatrix{1\cr
0}=\cr\hfill =f(x)\pmatrix{1\cr 0},\hfill\llap{(7)}\cr}$$ where
coefficient $C$ is equal to $(\frac{1}{d L_y \sqrt{\pi}})^{1/2}$,
$L_y$ is the size of the system in the $y$ direction. The variance
of the position operator $<(\Delta x)^2>$ in this case is equal to
$d^2/2$ and the variance $<(\Delta y)^2>$ exceed this value. The
variance of the momentum operator $p_x$ is $<(\Delta
p_x)^2>=\hbar^2/2d^2$, and the average ${\bf \hat p}$ is equal to
$p_{0x}$. One may consider the initial wave function as the
limiting case of a 2D packet with the width along $y$ direction
much greater than along $x$ i.e., $L_y\gg d$.

The electron wave function at any arbitrary moment of time can be
found with the help of the Green's function $$\pmatrix{\Psi_1(x,t)
\cr \Psi_2(x,t)}=\int dx^\prime dy^\prime
\pmatrix{G_{11}f(x^\prime)\cr G_{21}f(x^\prime)},\eqno(8)$$ where
matrix elements $G_{11}$ and $G_{21}$ of the matrix (5) are
determined by Eqs.(2),(3) and (6) $$\displaylines{G_{11}({\bf
r},{\bf r^\prime},t)=\frac{1}{(2\pi\hbar)^2}\int
\exp(-\frac{ip^2t}{2m\hbar}+i\frac{{\bf p}({\bf r}-{\bf
r^\prime})}{\hbar})\times\cr\hfill \times\cos(\frac{\alpha p
t}{\hbar})d{\bf p},\hfill\llap{(9)}\cr}$$

$$\displaylines{G_{21}({\bf r},{\bf
r^\prime},t)=\frac{1}{(2\pi\hbar)^2}\int
\exp(-\frac{ip^2t}{2m\hbar}+i\frac{{\bf p}({\bf r}-{\bf
r^\prime})}{\hbar})\times\cr\hfill \times\sin(\frac{\alpha p
t}{\hbar})\frac{p_x+ip_y}{p}d{\bf p}.\hfill\llap{(10)}\cr}$$ By
using the formula $$\displaylines{{\rm
e}^{iq\cos\psi}=J_0(q)+2\sum\limits_{n=1}J_{2n}(q)\cos(2n\psi)+\cr\hfill
+2i\sum\limits_{n=1}J_{2n-1}(q)\sin((2n-1)\psi).\hfill\llap{(11)}\cr}$$
and by integrating over the angle variable in Eqs.(9), (10) we
finally have
$$\displaylines{G_{11}=\frac{1}{2\pi\hbar^2}\int\limits_{0}^{\infty}\exp(-i\frac{p^2t}{2m\hbar}J_0(\frac{p|{\bf
r}-{\bf r^\prime}|}{\hbar})\times\cr\hfill\times\cos(\frac{\alpha
p t}{\hbar})p d p,\hfill\llap{(12)}\cr}$$

$$\displaylines{G_{21}=\frac{(x-x^\prime)+i(y-y^\prime)}{2\pi\hbar^2|{\bf
r}-{\bf
r^\prime}|}\int\limits_{0}^{\infty}\exp(-i\frac{p^2t}{2m\hbar}\times\cr\hfill\times
J_1(\frac{p|{\bf r}-{\bf r^\prime}|}{\hbar})\sin(\frac{\alpha p
t}{\hbar})p d p,\hfill\llap{(13)}\cr}$$ where $J_0$ and $J_1$ are
Bessel functions. Substituting Eqs. (12), (13) and (7) into Eq.(8)
and integrating over $x^\prime$ and $y^\prime$, we find the
analytical expression for the spinor components
$\psi_{1,2}(x,t)$.It should be noted that two electron bands with
chirality "plus" and "minus" give different contribution to the
electron wave functions. The calculation of the expressions for
$\Psi_{1,2}$ leads to the following electron probability densities
$|\Psi_1|^2$ and $|\Psi_2|^2$ at any arbitrary moment of the time
$$\displaylines{|\Psi_1|^2=\frac{C^2}{\sqrt{1+\gamma^2t^2}}\Bigg[\exp(-\frac{(x+(\alpha-\hbar
k_0/m )t)^2}{d^2(1+\gamma^2t^2)})+\cr\hfill
+\exp(-\frac{(x-(\alpha+\hbar k_0/m
)t)^2}{d^2(1+\gamma^2t^2)})+\hfill\cr \hfill
+2\exp\Bigg(-\frac{(x+(\alpha-\hbar k_0/m
)t)^2}{2d^2(1+\gamma^2t^2)}+\hfill\cr
\hfill+\frac{(x-(\alpha+\hbar k_0/m
)t)^2}{2d^2(1+\gamma^2t^2)}\Bigg)\times\hfill\cr\hfill
\times\cos(\frac{2(k_0 d^2+\gamma t x)\alpha
t}{d^2(1+\gamma^2t^2)})\Bigg],\hfill\llap{(14a)}\cr}$$

$$\displaylines{|\Psi_2|^2=\frac{C^2}{\sqrt{1+\gamma^2t^2}}\Bigg[\exp(-\frac{(x+(\alpha-\hbar
k_0/m )t)^2}{d^2(1+\gamma^2t^2)})+\cr\hfill
+\exp(-\frac{(x-(\alpha+\hbar k_0/m
)t)^2}{d^2(1+\gamma^2t^2)})-\hfill\cr \hfill
-2\exp\Bigg(-\frac{(x+(\alpha-\hbar k_0/m
)t)^2}{2d^2(1+\gamma^2t^2)}+\hfill\cr
\hfill+\frac{(x-(\alpha+\hbar k_0/m
)t)^2}{2d^2(1+\gamma^2t^2)}\Bigg)\times\hfill\cr \hfill
\times\cos(\frac{2(k_0 d^2+\gamma t x)\alpha
t}{d^2(1+\gamma^2t^2)})\Bigg],\hfill\llap{(14b)}\cr}$$ where
$\gamma=\hbar/d^2m$ is the inverse broadening time $p_{0x}=\hbar
k_0$.

As follows from Eqs.(14a), (14b) the shape of the function
$\rho(x,t)$ essentially depends on the parameter
$\eta=\frac{m^2\alpha^2d^2}{\hbar^2}$. In the case wide packet
when the momentum variance is much more $(m\alpha)^2$ and the
inequality $\eta\ll 1$ takes place, the evolution looks like at
the absence of Rashba term. Otherwise when $\eta\gg 1$ the initial
wave packet splits into two parts which propagate with different
group velocity, so the distance between these two parts increases
linear in time. This two parts correspond to the first and second
terms in square brackets in Eq.(14a) and Eq.(14b).The third terms
in Eq.(14a), (14b) describe the oscillation of the components of
electron density $|\Psi_1|^2$ and $|\Psi_2|^2$ in the region of
the overlapping of two split parts of the packet. It is clear that
these oscillations originates from the interference between the
states of different spectrum branches. When two parts of the
packet move away from each over the amplitude of the oscillations
decreases. The period of these oscillations along the $x$
direction depends on the initial width of the packet $d$ and
equals to $\Delta x=\pi d^2(1+\gamma^2t^2)/\alpha\gamma t^2$. So,
if inequality $\gamma t\ll 1$ takes place the period of
oscillation decreases with time and equals to $\Delta x=\pi m
d^4/\alpha \hbar t^2$ and when $\gamma t\gg 1$ the oscillation
period is not depend on the time $\Delta x=\pi \hbar/m\alpha$.

To illustrate the evolution of the electron probability density
$\rho(x,y)=|\Psi_1|^2+|\Psi_2|^2$ we plot this function using
Eq.(14a), (14b) at the Fig.1(a) for the moments of the time: $t=0,
t=1,5, t=7$ (in the units of $\tau_0=\gamma^{-1}$). The
calculations was made for the parameters $GaAs/InGaAs$ electron
system: $m=0,05m_0,\, \alpha=3,6\cdot 10^6\, {\rm cm}\cdot {\rm
sec^{-1}}$ and the packet parameters: $d=10^{-5}\, {\rm cm},\,
k_0=2,5\cdot 10^5\, {\rm cm^{-1}}$. Here one can clearly see that
initial Gaussian wave packet Eq.(7) splits up at $t>0$ into two
parts propagating along the $x$ direction. The width of each part
of the packet increases in time as for the case of free particle.

To analyze spin dynamics one can consider the time evolution of
the spin density
$$s_i(x,y,t)=\frac{\hbar}{2}(\Psi_1^\ast,\Psi_2^\ast)\hat \sigma_i
\pmatrix{\Psi_1\cr \Psi_2},\eqno(15)$$ Using Eqs.(14a) and (14b)
we immediately find the expression for spin density
$s_z=\frac{\hbar}{2}(|\Psi_1({\bf r},t)|^2-|\Psi_2({\bf
r},t)|^2)$, which demonstrate the oscillatory behavior as a
function of $x$ (see Fig.1(b)). The period of oscillation here is
the same as for the functions $|\Psi_{1,2}(x,t)|$. For the spin
density $s_y(x,t)$ the following result can be obtained
$$\displaylines{s_y(x,t)=\frac{\hbar}{\sqrt{\pi}L_y d
\sqrt{1+\gamma^2 t^2}}\times\cr\hfill \Bigg[\exp(-\frac{(x-(\hbar
k_0/m-\alpha )t)^2}{d^2(1+\gamma^2t^2)})-\hfill\cr\hfill
-\exp(-\frac{(x-(\hbar k_0/m+\alpha
)t)^2}{d^2(1+\gamma^2t^2)})\Bigg],\hfill\llap{(16)}\cr}$$

According to this equation both pats of the initial wave packet
moving along the $x$ direction with different velocities are
characterized by the opposite spin orientation (at the same time
the average spin component $\bar S_y=\int s_y(x,t)d{\bf r}$ is
equal to zero).

Note that the components of wave function depend only on
coordinate $x$, that leads to $\bar p_y=p_y=0$, however the
velocity $\bar v_y(t)\neq 0$. Really, using the definition Eq.(4)
it is not difficult to obtain $$\displaylines{\bar
v_x(t)=\frac{\hbar k_0}{m},\cr\hfill \bar v_y(t)=-\alpha\sin(2k_0
\alpha t)\exp(-(\frac{\alpha t}{d})^2).\hfill\llap{(17)}\cr}$$ As
follows from these equations the average $\bar v_y$ velocity
performs the oscillations in the transverse direction ({\it
zitterbewegung} or jittering) with the frequency $2k_0\alpha$ and
the damping time is determined by the parameter $d/\alpha$.

\section{Evolution of two dimensional packets at the presence of spin orbit coupling}
We consider now the evolution of two dimensional wave packet at
the presence of spin orbit coupling. Let us consider the following
form of the Gaussian packet at the initial moment $t=0$:
$$\displaylines{\Psi({\bf
r},0)=C\exp(-\frac{r^2}{2d^2}+ip_{0x}x/\hbar)\pmatrix{1\cr
0}=\cr\hfill =f({\bf r})\pmatrix{1\cr 0},\hfill\llap{(18)}\cr}$$
where $p_{0x}=\hbar k_0$ is the average momentum and
$C=1/\sqrt{\pi}d$. Then, using a Green's function method we arrive
after some algebra at the following equations for the components
of spinor (in  the momentum space) $$\displaylines{C_1({\bf
p},t)=\frac{d}{\sqrt{\pi}\hbar}\cos(\frac{\alpha p
t}{\hbar})\exp(-\frac{ip^2t}{2m\hbar}-\cr\hfill-\frac{p^2d^2}{2\hbar^2}-\frac{k_0^2d^2}{2}+\frac{p_x
k_0 d^2 }{\hbar}),\hfill\llap{(19a)}\cr}$$

$$\displaylines{C_2({\bf
p},t)=-\frac{d}{\sqrt{\pi}\hbar}\frac{p_x+ip_y}{p}\sin(\frac{\alpha
pt}{\hbar})\exp(-\frac{ip^2t}{2m\hbar}-\cr\hfill
-\frac{p^2d^2}{2\hbar^2}-\frac{k_0^2d^2}{2}+\frac{p_x k_0 d^2
}{\hbar}),\hfill\llap{(19b)}\cr}$$

After that $\Psi_{1,2}({\bf r},t)$ can be obtained directly by 2D
Fourier transform of $C_{1,2}({\bf r},t)$: $$\displaylines{
\Psi_1({\bf
r},t)=\frac{d}{\sqrt{\pi}}\int\limits_0^\infty\exp(-i\frac{q^2\hbar
t}{2m}-\frac{q^2d^2}{2}-\frac{k_0^2d^2}{2})\times\cr \hfill \times
I_0(q\sqrt{k_0^2d^4-r^2+2ik_0d^2r\cos\varphi})\times\hfill\cr
\hfill \times\cos(\alpha qt)qdq,\hfill\llap{(20a)}\cr}$$

$$\displaylines{\Psi_2({\bf
r},t)=-\frac{id}{\sqrt{\pi}}\frac{r\cos\varphi+ir\sin\varphi-ik_0d^2}{\sqrt{r^2-k_0^2d^4-2ik_0d^2r\cos\varphi}}
\times\cr \hfill \times\int\limits_0^\infty\exp(-i\frac{q^2\hbar
t}{2m}-\frac{q^2d^2}{2}-\frac{k_0^2d^2}{2})\times\hfill \cr \hfill
\times J_1(\sqrt{(r^2-k_0^2d^4-2ik_0d^2r\cos\varphi})\times\hfill
\cr \hfill \times\sin(\alpha qt)qdq,\hfill\llap{(20b)}\cr}$$ where
$J_1$ and $I_0$ are the Bessel and the modified Bessel functions
of the first and the zeroth order, $\varphi$ is asimutal angle in
the $xy$ plane. These expressions become simpler if the average
momentum of a wave packet is equal to zero, i.e. $p_{0x}=0$. In
this case $$
\displaylines{\Psi_1=\frac{d}{\sqrt{\pi}}\int\limits_0^\infty
qJ_0(qr)\cos(\alpha qt)\times\cr \hfill
\times\exp(-i\frac{q^2\hbar
t}{2m}-\frac{q^2d^2}{2})dq,\hfill\llap{(21a)}\cr}$$

$$\displaylines{\Psi_2=\frac{d}{\sqrt{\pi}}\frac{y-ix}{r}\int\limits_0^\infty
qJ_1(qr)\sin(\alpha qt)\times\cr \hfill
\times\exp(-i\frac{q^2\hbar
t}{2m}-\frac{q^2d^2}{2})dq.\hfill\llap{(21b)}\cr}$$

As in the case of 1D packet the shape of the full electron density
$\rho(x,t)=|\Psi_1|^2+|\Psi_2|^2$ at $t>0$ depends on the
parameter $\eta=\frac{m^2\alpha^2 d^2}{\hbar^2}$. In  Fig. 2. we
show the electron density $\rho(x,t)$ for the case $p_{0x}=0$ at
the time $t=5$ (in the units of $\frac{d}{\alpha}$) and
$\eta=2,7$. As one can see the spin-orbit coupling qualitatively
change the character of the wave packet evolution, so that during
the time the initial Gaussian packet turns into two axially
symmetric parts. As follows from our analytical and numerical
calculation the outer part propagates with group velocity which is
greater than $\alpha$   and the inner part moves with group
velocity lower than  $\alpha$. If $\eta\ll 1$, i.e. the packet is
narrow enough, its evolution remained the standard broadening of
the Gaussian packet of free particle.

In  Fig. 3(a) it is shown the packet evolution for the case $\bar
p_{0x}=\hbar k_0\neq0$. It is clear that in this case the
cylindrical symmetry is absent, and two maximums of the electron
density spread along the $x$ direction with not equal velocities.
Each one of these two parts are spin polarized. Fig. 3(b)
illustrates the distribution of the spin polarization
$s_y(x,y,t)$for the initial state, polarized along $z$ axis,
Eq.(18). It is a smooth function which has different sign in the
regions for two maximums of the electron density.

When $\bar p_x\neq0$ the motion of the wave packet center along
$x$ accompanied  by the oscillation of the packet center in a
perpendicular direction, or {\it zitterbewegung}. Below we
consider the effect of damping of {\it zitterbewegung} oscillation
for 2D packet which was not predicted in\cite{Chen}.

Using Eq.(22a) and Eq.(22b) we calculate the average value of the
operator $\hat y=i\hbar\frac{\partial}{\partial p_y}$ and obtain
for $t>0$ the result $$\displaylines{\bar
y(t)=-\frac{2d^2}{\hbar}\exp[-(k_0d)^2]\int\limits_0^\infty\sin^2(\frac{\alpha
pt}{\hbar})\times\cr \hfill
\times\exp(-\frac{p^2d^2}{\hbar^2})I_1(\frac{2pk_0d^2}{\hbar})dp.\hfill\llap{(22)}\cr}$$

In the case when wave packet is wide enough and the inequality
$a=dk_0\gg 1$ takes place, one can obtain a simple asymptotic
formula for $\bar y(t)$. To show this we represent Eq. (22) as a
sum of two terms $$\displaylines{\bar
y(t)=-d\exp(-a^2)\Bigg[\int\limits_0^\infty\exp(-u^2)I_1(2au)du-\cr\hfill
-\int\limits_0^\infty\cos(\frac{2\alpha t u
}{d})\exp(-u^2)I_1(2au)du\Bigg]\hfill\cr
\hfill=-d\exp(-a^2)\Bigg[\frac{1}{2a}(\exp(a^2)-1)-Z\Bigg],\hfill\llap{(23)}\cr}$$
where we denote $\frac{pd}{\hbar}=u$,
$Z=Re(\int\limits_0^\infty\exp(-u^2+i\frac{2\alpha t
u}{d})I_1(2au)du)$. To evaluate $Z$ we replace the modified Bessel
function $I_1(2dk_0u)$ by it's asymptotic formula
$I_1(x)=\frac{{\rm e}^x}{\sqrt{2\pi}x}$, which valid for the case
$k_0d\gg 1$. After that the integral with respect to $u$ can be
evaluated using the stationary phase method that leads to the
simple result $$
Z=\frac{1}{2kd}\exp(a^2-\frac{\alpha^2t^2}{d^2})\cos(2\alpha
k_0t).$$ Substituting this expression into Eq.(23) we finally have
$$ \bar
y(t)=-\frac{1}{2k_0}\Bigg[1-\exp(-\frac{\alpha^2t^2}{d^2})\cos(2\alpha
k_0 t)\Bigg].\eqno(24)$$ The last result demonstrates clearly that
$\bar y(t)$ experience the damped oscillations with the frequency
$2\alpha k_0$ decaying for the time  $\frac{d}{\alpha}$. In the
real 2D structures the frequency of the {\it zitterbewegung} have
the order of $10^{11}-10^{12}\,{\rm sec^{-1}}$ for $k_0\approx
10^{-5}-10^{-6}\/ {\rm cm}$. The amplitude of the {\it
zitterbewegung} is proportional to  the  electron wavelength  in
$x$. At Fig. 4 we plot the function $\bar y(t)$ determined by Eq.
(22) which demonstrates in accordance with Eq.(24) the effect of
{\it zitterbewegung} damping. When $t\gg\frac{d}{\alpha}$ the
oscillations stop and the center of the wave packet is shifts in
direction perpendicular to the group velocity at the value of
$1/2k_0$. The last result coincides with\cite{Chen}.  Since the
packet moves with constant velocity, the time oscillations of
$\bar y(t)$ can be easily converted to the oscillation of the wave
packet center in real $x,\,y$ space.

\section{Cyclotron dynamics of 2D wave packet in a perpendicular magnetic field}
In this section we examine the cyclotron dynamics of electron wave
packet rotating in a  magnetic  field ${\bf B}(0,0,B)$ which is
perpendicular to the plane of 2D electron gas.  In this case the
one-electron Hamiltonian including Rashba term reads
$$\displaylines{H=\frac{({\bf \hat p}+e{\bf
A}/c)^2}{2m}+\alpha(\hat \sigma_y (\hat p_x+eA_y/c)-\hat \sigma_x
\hat p_y)+\cr\hfill +g\mu_B\sigma_z.\hfill\llap{(25)}\cr}$$ Here
$e$ is the electron charge, $m$ is the effective mass, $\hat
p_{x,y}$ are the momentum operator components, $\alpha$ is the
parameter of Rashba coupling, $g$ is the Zeeman factor, and
$\mu_B$ is the Bohr magneton. Bellow we use the Landau gauge for
the vector potential ${\bf A}=(-By,0,0)$. Then the eigenvalues and
the eigenfunctions of the Hamiltonian (25) indicating by quantum
numbers $n,\, k_x,\, s=\pm 1$ and corresponding to two branches of
levels  can be evaluated analytically (see,
e.g.,\cite{Wan})$$E_n^{\pm}=\hbar
\omega_cn\pm\Bigg(E_0^2+\frac{2n\alpha^2\hbar^2}{\ell_B^2}\Bigg)^{1/2},\eqno(26)$$
where $E_0^+=\frac{\hbar \omega_c}{2}-g\mu_BB$ is the zero Landau
level, $n=1,2,3,\ldots$, $\omega_c=\frac{eB}{mc}$ is the cyclotron
frequency, $\ell_B=\sqrt{\frac{\hbar}{m\omega_c}}$ is the magnetic
length. The  eigenspiniors  are
$$\displaylines{\psi_{n,k_x}^+({\bf r})=\frac{{\rm
e}^{ik_xx}}{\sqrt{2\pi A_n}}\pmatrix{-iD_n\phi_{n-1}(y-y_c)\cr
\phi_n(y-y_c)},\cr\psi^-_{n,s}=\frac{{\rm e}^{ik_xx}}{\sqrt{2\pi
A_n}}\pmatrix{\phi_{n-1}(y-y_c)\cr -iD_n\phi_s(y-y_c)},\cr \hfill
\psi_0^+=\frac{{\rm e}^{ik_xx}}{\sqrt{2\pi}}\pmatrix{0\cr
\phi_0(y-y_c)}.\hfill\llap{(27)}\cr}$$ Here coefficients $D_n$ are
given by:
$D_n=\frac{\sqrt{2n}\alpha\hbar/\ell_B}{E_0+\sqrt{E_0^2+2n\alpha^2\hbar^2/\ell_B^2}},\,
A_n=1+D_n$, $\phi_m(y-y_c)$ are linear oscillator wave functions,
$y_c=\ell_B^2k_x$ is the center of oscillator. It should be noted
that for enough weak magnetic field the dependence of energy
$E_n^-$ on quantum number $n$ $(n\gg 1)$ resembles the behavior of
the function $\varepsilon_-(p)$, Eq. (3). Namely, for small n the
values of energy $E_n^+$ are negative, decreasing with $n$, like
for the hole states.

Using the Eqs.(26),(27) we can obtain components of the matrix
Green's function which permits us to find the time evolution of
the initial state. The usual definition
$$\displaylines{G_{ij}({\bf r},{\bf r
}^\prime,t)=\sum\limits_{s=\pm}\int dk_x\times\cr\hfill
\times\sum\limits_{n=0}^\infty\psi_{n,k_x,i}^s({\bf
r},t)\psi_{n,k_x,j}^{\ast s}({\bf r},0).\hfill\llap{(28)}\cr}$$
yields $$\displaylines{G_{11}({\bf r},{\bf
r}^\prime,t)=\frac{1}{2\pi}\int\limits_{-\infty}^{+\infty}dk_x{\rm
e}^{ik_x(x-x\prime)}\times\cr\hfill
\times\sum\limits_{n=0}f_{n+1}(t)\phi_n(y-y_c)\phi_n(y\prime-y_c),\hfill\llap{(29a)}\cr}$$

$$\displaylines{G_{21}({\bf r},{\bf
r}^\prime,t)=\frac{1}{2\pi}\int\limits_{-\infty}^{+\infty}dk_x{\rm
e}^{ik_x(x-x\prime)}\times\cr\hfill
\times\sum\limits_{n=0}g_{n+1}(t)\phi_{n+1}(y-y_c)\phi_n(y\prime-y_c),\hfill\llap{(29b)}\cr}
$$ where the time-dependent coefficients $f_n(t)$ and $g_n(t)$ are
given by $$f_n(t)={\rm e}^{-i\omega_c
nt}(\cos\delta_nt-i(1-\frac{2}{A_n})\sin\delta_nt),\eqno(30a)$$
$$g_n(t)={\rm e}^{-i\omega_c
nt}\frac{2D_n}{A_n}\sin\delta_nt,\eqno(30b)$$ and
$\delta_n=\frac{1}{\hbar}\sqrt{E_0^2+\frac{2n\alpha^2\hbar^2}{\ell_B^2}}.$

Let the initial state coincides with the wave function of the
coherent state in a magnetic field $$\displaylines{\Psi({\bf
r},0)=\frac{1}{\sqrt{\pi}\ell_B^2}\exp(-\frac{r^2}{2\ell_B^2}+ip_{0x}x/\hbar)\pmatrix{1\cr
0}.\hfill\llap{(31)}\cr}$$ Such choice of wave function $\Psi({\bf
r},0)$ is motivated by the following: as it is well known, at the
absence of spin orbit coupling the dynamics of coherent states in
a magnetic field looks like the dynamics of a classical particle.
To analyze the time evolution in our case one needs to calculate
the wave function at $t>0$. Straightforward algebra with using
Eqs.(29a), (29b), (31) leads to the final expressions
$$\displaylines{\psi_1({\bf
r},t)=\frac{1}{\sqrt{2}\pi\ell_B}\sum\limits_{n=0}\frac{f_{n+1}(t)}{2^nn!}\int\limits_{-\infty}^{+\infty}du{\rm
e}^{\varphi(x,y,u)}\times\cr\hfill
\times(-u)^nH_n(\frac{y}{\ell_B}-u),\hfill\llap{(32a)}\cr}$$

$$\displaylines{\psi_2({\bf
r},t)=\frac{1}{2\pi\ell_B}\sum\limits_{n=0}\frac{g_{n+1}(t)}{2^nn!\sqrt{n+1}}\int\limits_{-\infty}^{+\infty}du{\rm
e}^{\varphi(x,y,u)}\times\cr\hfill
\times(-u)^nH_{n+1}(\frac{y}{\ell_B}-u),\hfill\llap{(32b)}\cr}$$
where
$\varphi(x,y,u)=iu\frac{x}{\ell_B}-\frac{(p_{0x}\ell_B/\hbar-u)^2}{2}-\frac{u^2}{4}-\frac{y-u\ell_B}{2\ell_B}.$

The electron density obtained by numerical evaluations of the
integrals Eqs.(32a) and (32b) is represented in Fig. 5 for
relatively weak spin orbit coupling and strong magnetic field. The
calculations was made for the material parameters of two
dimensional $GaAs$ heterostructure: $m=0,067m_0,\,
\alpha=3,6\cdot10^5\, cm\cdot sec^{-1},\, g=-0,44,\, B=1T$ and
$k_{0x}=p_{0x}/\hbar=1,5\cdot 10^6\, cm^{-1}.$ It is not difficult
to verify that the series in Eqs. (32a) and (32b) converge very
rapidly as $n$ increases. So for our parameters it suffices to
take $n_{max}=25$ to calculate the components $\psi_1({\bf r},t)$
and $\psi_2({\bf r},t)$. At $t>0$ the initial wave packet (Fig.
5(a)) splits on two parts (Fig. 5(b)) which "rotate" with
different incommensurable cyclotron frequencies. In accordance
with Eqs.(26) these  frequencies can be determine by the
expressions
$$\displaylines{\omega_c^\pm=\frac{E_{n+1}^\pm-E_n^\pm}{\hbar}=\omega_c\pm\sqrt{E_0^2+
2(n+1)\frac{\alpha^2\hbar^2}{\ell_B^2}}\mp\cr\hfill\mp\sqrt{E_0^2+2n\frac{\alpha^2\hbar^2}
{\ell_B^2}}.\hfill\llap{(33)}\cr}$$ The effective $n$ in this
equation is connected with cyclotron radius via relation
$R_c(t)=\frac{p_{0x}}{m\omega_c}=\sqrt{2n}\ell_B$. Believing that
$\varsigma=\frac{2n\alpha^2\hbar^2}{\ell_B^2E_0^2}\ll 1$, i.e. in
the case of a weak spin orbit coupling or strong magnetic field,
one can obtain from (33) the approximate expression for the
difference between cyclotron frequencies
$$\omega_c^+-\omega_c^-=2\frac{\alpha^2m}{E_0}\omega_c.\eqno(34)$$
Fig.5(b) demonstrates for the case $\varsigma\ll 1$   the
distribution of electron density at the moment when two parts are
located at opposite points of the cyclotron orbit. The
correspondent time can be determined from the relation
$(\omega_c^+-\omega_c^-)t_0=\pi$ and hence for the $GaAs$
structure we will have
$t_0=\frac{\pi}{\omega_c^+-\omega_c^-}=\frac{\pi}{\omega_c}\frac{E_0}{2\alpha^2m}=45\frac{2\pi}{\omega_c}$.

After some cyclotron periods two split packets merge again which
is demonstrated in Fig.5(c). With time due to the effect of the
broadening electron probability distributes randomly around
cyclotron orbit that is shown in Fig.5(d).

In the opposite case of relatively strong spin orbit coupling or
weak magnetic field when the inequality
$\varsigma=\frac{2n\alpha^2\hbar^2}{\ell_B^2E_0^2}\gg 1$ holds
true the difference between two cyclotron frequencies, as it
follows from Eq.(33), equals to
$\omega_c^+-\omega_c^-=\frac{\sqrt{2}\alpha}{\sqrt{n}\ell_B}$. For
the $InGaAs$ structure with parameters $m=0,05m_0,\,
\alpha=3,6\cdot10^6\, cm\cdot sec^{-1},\, g=-10,\, B=1T$ and
$k_{0x}=p_{0x}/\hbar=1,5\cdot 10^6\, cm^{-1}$, we have
$\varsigma=8$ and the divergence time $t_0\approx 2,3\cdot
\frac{2\pi}{\omega_c}$.

One can analyze the effects of the periodic splitting and
reshaping of the wave packet in magnetic field as well as the
process of distribution around cyclotron orbit by considering the
time dependence of the cyclotron radius determined as
$R(t)=\sqrt{\{\bar x(t)\}^2+\{\bar y(t)\}^2}$. To do this we
represent the average value of coordinates $x_1=x$, $x_2=y$ as
$$\displaylines{\bar x_i=\int\psi_1^\ast({\bf r},t)x_i\psi_1({\bf
r},t)d{\bf r}+\cr\hfill+\int\psi_2^\ast({\bf r},t)x_i\psi_2({\bf
r},t)d{\bf r},\hfill\llap{(35)}\cr}$$ where $\psi_1$ and $\psi_2$
are determined by Eqs. (32a) and (32b). The lengthy calculations
(see the Appendix) eventually yield the explicit expression for
the $\bar x(t)$ and $\bar y(t)$: $$\displaylines{\bar
x(t)=-\frac{\ell_B}{3}\exp(-\frac{p_{0x}^2\ell_B^2}{3\hbar^2})\times\cr\hfill\times\{\cos\omega_ct\sum\limits_{k=0}S_k(t)H_{2k+1}(i\sqrt{\frac{2}{3}}p_{0x}
\ell_B/\hbar)+\hfill\cr\hfill+\sin\omega_ct\sum\limits_{k=0}P_k(t)
H_{2k+1}(i\sqrt{\frac{2}{3}}p_{0x}\ell_B/\hbar)\},~~~\hfill\llap{(36a)}\cr}$$

$$\displaylines{\bar
y(t)=q\ell_B^2+\frac{\ell_B}{3}\exp(-\frac{p_{0x}^2\ell_B^2}{3\hbar^2})\times\cr\hfill\times
\{\cos\omega_ct\sum\limits_{k=0}P_k(t)H_{2k+1}(i\sqrt{\frac{2}{3}}p_{0x}\ell_B/\hbar)-\hfill\cr\hfill
-\sin\omega_ct\sum\limits_{k=0}S_k(t)H_{2k+1}(i\sqrt{\frac{2}{3}}p_{0x}\ell_B/\hbar)\}.~~~~\hfill\llap{(36b)}\cr}$$
As one can see, the dependence of  $\bar x(t)$ and $\bar y(t)$ on
the time are determined by both the factors $\cos \omega_c t$ and
$\sin \omega_c t$ as well as by functions
$$\displaylines{S_k(t)=i\frac{(-1)^k}{k!}(\frac{1}{12})^k[\xi_{k+2}\cos\delta_{k+1}t\sin\delta_{k+2}t-\cr\hfill
-\xi_{k+1}\cos\delta_{k+2}t\sin\delta_{k+1}],\hfill\llap{(37a)}\cr}$$

$$\displaylines{P_k(t)=i\frac{(-1)^k}{k!}(\frac{1}{12})^k[\cos\delta_{k+1}t\cos\delta_{k+2}t+\cr\hfill+
(\xi_{k+1}\xi_{k+2}+
4\sqrt{\frac{k+2}{k+1}}\frac{D_{k+1}D_{k+2}}{A_{k+1}
A_{k+2}})\times\hfill\cr\hfill\times\sin\delta_{k+1}t\cos\delta_{k+2}t],\hfill\llap{(37b)}\cr}$$
$$\xi_k=\frac{D_k^2-1}{D_k^2+1},$$ which describe the additional
time dependence due to spin precession. Note that the frequencies
$\delta_k=\frac{1}{\hbar}\sqrt{E_0^2+\frac{2k\alpha^2\hbar^2}{\ell_B^2}}$
are incommensurable. As a check on this formalism, it is not
difficult to show that in the absence of Rashba coupling
$(\alpha=0)$ as it follows from Eqs. (36a) and (36b)
$$\displaylines{\bar x(t)=p_{0x}\ell_B^2/\hbar\sin\omega_c
t,\cr\hfill \bar y(t)=p_{0x}\ell_B^2(1-\cos\omega_c
t),\hfill\llap{(38)}\cr}$$ that correspond to the classical motion
of charged particle in the magnetic field with a constant radius.

The time dependence of the cyclotron radius $R(t)$ in the system
with Rashba is presented at Fig. 6. It is clear that the
oscillations of $R(t)$ are connected with the effects of periodic
splitting and reshaping of wave packets.  The radius has the
minimal values at the moments when two parts of the packet are
located at the opposite point of cyclotron orbit. This situation
is shown at FIG 5.(b). The first minimum labeled by the letter
${\bf {\it b}}$ at FIG .6   One can see that the time of the first
minimum  approximately coincides to our estimation made above:
$t_0\approx 45T_c$. The radius is maximal at the moments of the
packet reshaping that is shown at Fig. 5(c) (two of these points
labeled by the letters ${\bf {\it a}}$ and ${\bf {\it c}}$. Due to
the effects of incommensurability of the cyclotron frequencies and
the packet broadening the amplitude of the oscillations decrees
with the time After that the electron density distributes around
cyclotron orbit, the amplitude of the oscillation ceases and the
electron density distribution acquire the no regular character
(Fig.5(d)).

We evaluate also the distribution of the electron density for the
structure with relatively strong spin orbit coupling. For such
systems instead of the repeated process of the splitting and
restoring of the wave packet discussed above the transition to the
irregular distribution along the cyclotron orbit can be realized
for the time of the order of one cyclotron period. This conclusion
is confirmede by the simple estimation made for the $InGaAs/GaAs$
structure discussed above.

\section{Conclusions}

We have analyzed the evolution of 1D and 2D wave packets in 2D
electron gas with linear Rashba spin orbit coupling. We showed
that the electron packet dynamics differs drastically from usual
quantum dynamics of electrons with parabolic energy spectrum.
Depending on the  initial spin polarization packet splits in two
parts which propagate with different velocities and have different
spin orientation. At the time when two parts of wave packet
overlap, the packet center performs oscillations in much the same
way as for a relativistic particle. The direction of these
oscillations is perpendicular to the packet group velocity. When
the distance between split parts exceeds the initial width of the
packet these oscillations stop.

In the 2D semiconductor structures placed in a perpendicular
magnetic field the spin orbit coupling changes the cyclotron
dynamics of charged particles. As at the absence of magnetic field
the initial packet splits in two parts, which rotate in a
perpendicular magnetic field with different incommensurable
cyclotron frequencies.  As a result, after some cyclotron periods
these parts join again. The corresponding time $t_0$ essentially
depends upon the ratio of the energy of spin orbit coupling and
the distance between Landau levels, Eq.(26):
$\varsigma=\frac{2n\alpha^2\hbar^2}{\ell_B^2E_0^2}$. Thus, for the
systems with weak and relatively strong spin orbit coupling e.g.
$GaAs$ and $InGaAs$ hetrodtructures,  the time $t_0$ equals to
$45T_c$ and $2,3T_c$, respectively.

The splitting and {\it zitterbewegung} of the wave packets in
nanostructures with spin-orbit coupling can be observed
experimentally in low dimensional structures. In particular, these
effects should determine the electron dynamics and high-frequency
characteristics of the field effect transistor by Datta and
Das\cite{Wan}, and other spintronic devices. Simple estimations
show that during the time of the wave packet propagation through
the ballistic transistor channel where the distance between
emitter and collector is of the order of $1\mu m$, the distance
between two split parts of the wave packet becomes comparable with
its initial size.  In this situation the high-frequency
characteristics of the field effect transistor should be
substantially affected by the spin-orbit coupling. Moreover, the
atypical semiclassical dynamics of a spin-orbit system placed in a
magnetic field will influence the shape of the cyclotron resonance
line in 2D systems with spin orbit coupling. An important feature
of these experiments is that the electron transport is in the
ballistic regime and thus the momentum relaxation time $\tau$
should be considered much more greater compared with the typical
splitting time.

\section*{Acknowledgments}
The authors are grateful to D.V. Khomitsky for useful discussions.
This work was supported by the program of the Russian Ministry of
Education and Science "Development of scientific potential of high
education" (project 2.1.1.2363).

\section{Appendix}

This appendix provides some of details involved in obtaining the
average value of the position operator given by Eqs.(36a), (36b).
According to Eq.(35) $$\bar y(t)=\bar y_1(t)+\bar
y_2(t).\eqno(A.1)$$ Consider the calculation of the first term
$\bar y_1(t)$. Using the initial wave function, Eq.(31), we obtain
$$\displaylines{\bar y_1(t)=\int\int\frac{d{\bf r^\prime}d{\bf
r^{\prime\prime}}}{\pi\ell_B^2}\times\cr\hfill\times\exp\Bigg(-\frac{r^{\prime2}+r^{\prime\prime2}}{2\ell_B^2}
+i\frac{p_{0x}(x^\prime-x^{\prime\prime})}{\hbar}\Bigg)\times\hfill\cr\hfill\times\int
G_{11}({\bf r},{\bf r^\prime},t)y G_{11}^\ast({\bf r},{\bf
r^{\prime\prime}},t)d{\bf r}.\hfill\llap{(A.2)}\cr}$$

Denote the last integral in this equation as
$$M_{11}^y=\int
G_{11}({\bf r},{\bf r^\prime},t) y G_{11}^\ast({\bf r },{\bf
r^{\prime\prime}},t)d{\bf r}.\eqno(A.3)$$

Then substituting Eq.(29a) into Eq.(A.3) and using the well known
formula for a linear harmonic oscillator functions
$$\displaylines{\int\limits_{-\infty}^{+\infty}y\phi_n(y-y_c)\phi_k(y-y_c)dy=\cr\hfill=\frac{\ell_B}{\sqrt{2}}\Bigg(\sqrt{n}\delta_{k,n-1}
+\sqrt{n+1}\delta_{k,n+1}\Bigg)+
y_c\delta_{n,k},~~~~~~~~~~\hfill\llap{(A.4)}\cr}$$ we will have
$$M_{11}^y=\frac{1}{2\pi}\int \limits_{-\infty}^{+\infty}{\rm
e}^{ik_x(x-x^{\prime\prime})\mu(y^\prime,y^{\prime\prime},t,y_c)}dk_x.\eqno(A.5)$$
Here $$\displaylines{\mu(y^\prime,y^{\prime\prime},t,y_c)=
\frac{\ell_B}{\sqrt{2}}\times\cr\hfill\times\Bigg(\sum\limits_{n=0}\sqrt{n}f_{n+1}(t)f_n^\ast(t)\phi_n(y^\prime-y_c)
\phi_{n-1}(y^{\prime\prime}-y_c)+\hfill\cr\hfill
+\sum\limits_{n=0}\sqrt{n+1}f_{n+1}(t)f_{n+2}^\ast(t)\times\hfill\cr\hfill\times\phi_n(y^\prime-y_c)
\phi_{n+1}(y^{\prime\prime}-y_c)\Bigg)+\hfill\cr\hfill
+y_c\sum\limits_{n=0}|f_{n+1}(t)|^2\phi_n(y^\prime-y_c)\phi_n(y-y_c),~~~~~~\hfill\llap{(A.6)}\cr}$$
where the coefficients $f_n(t)$ are given by Eq.(30a). We
calculate $\bar y_1(t)$ by substituting Eqs.(A.5), (A.6) into
(A.2). The resulting integrals can be evaluated by using Gaussian
transformation\cite{Gra}
$$\displaylines{\frac{1}{\sqrt{\pi}}\int\limits_{-\infty}^{+\infty}{\rm
e}^{-(x-y)^2}H_n(y)dy=(2x)^n,\cr\hfill
\frac{1}{\sqrt{\pi}}\int\limits_{-\infty}^{+\infty}{\rm
e}^{-(x-y)^2}y^ndy=\frac{H_n(ix)}{(2i)^n}.\hfill\llap{(A.7)}\cr}$$
Finally we obtain $$\displaylines{\bar y_1(t)=\frac{\ell_B}{6}\exp
(-\frac{(p_{0x}\ell_B)^2}{3\hbar^2})\times\cr\hfill
\times\sum\limits_{k=0}\psi_k(t)H_{2k+1}(i\sqrt{2/3}p_{0x}\ell_B/\hbar),\hfill\llap{(A.8)}\cr}$$
where
$$\displaylines{\psi_k(t)=\frac{i}{k!}\Bigg(-\frac{1}{12}\Bigg)^k(f_{k+1}^\ast(t)f_{k+2}(t)+\cr\hfill
+f_{k+1}(t)f_{k+2}^\ast(t)-2|f_{k+1}(t)|^2).\hfill\llap{(A.9)}\cr}$$
Performing the same kind of calculation we have for $\bar y_2(t)$:
$$\displaylines{\bar
y_2(t)=\frac{\ell_B}{6}\exp(-\frac{(p_{0x}\ell_B)^2}{3\hbar^2})\times\cr\hfill\times
\sum\limits_{k=0}\gamma_k(t)H_{2k+1}(i\sqrt{2/3}p_{0x}\ell_B/\hbar),~~\hfill\llap{(A.10)}\cr}$$
where
$$\displaylines{\gamma_k(t)=\frac{i}{k!}\Bigg(-\frac{1}{12}\Bigg)^k(g_{k+1}^\ast(t)g_{k+2}(t)+\cr\hfill
+g_{k+1}(t)g_{k+2}^\ast(t)-2|g_{k+1}|^2),\hfill\llap{(A.11)}\cr}$$
and the coefficients $f_k(t)$ and $g_k(t)$ in Eqs.(A.9), (A.11)
are determined by Eqs.(30a), (30b). The preceding expressions
immediately lead to the average value $\bar y(t)$ given in Eq.
(36b). The evaluation of $\bar x(t)$ can be obtained by following
the procedure similar to that which led to Eq.(36a).

\newpage
\begin{figure}
  \centering
  \includegraphics[width=85mm]{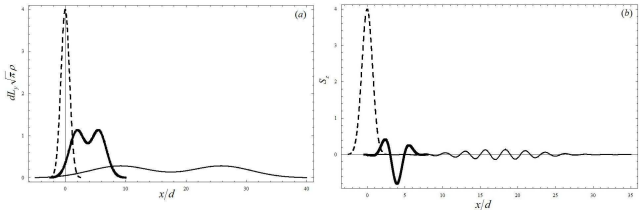}
\caption{The electron probability density
$\rho(x,t)=|\Psi_1|^2+|\Psi_2|^2(a),$ and spin density $s_z$(b).
The dashed, thick and thin lines correspond to different moments
of the time namely $t_1=0,t_2=1,5$ and $t_3=7$(in the units
$\tau_0=\gamma^{-1}$).} \label{fig1}
\end{figure}

\newpage

\begin{figure}
  \centering
  \includegraphics[width=85mm]{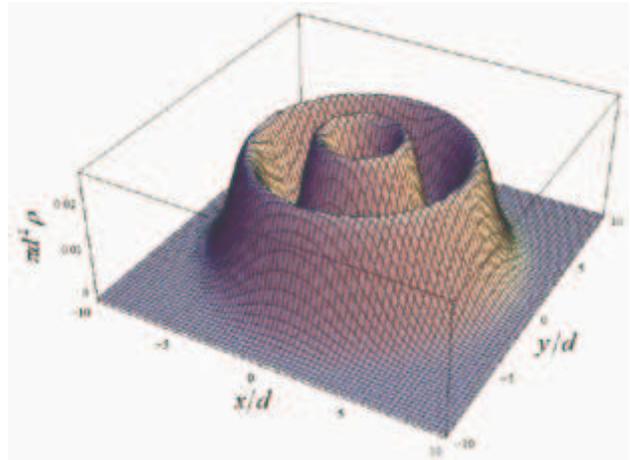}
\caption{(Color on line). The electron probability density
$\rho(x,t)=|\Psi_1|^2+|\Psi_2|^2$ for the initial state Gaussian
packet Eq.(18) with $p_{0x}=0$ at the time $t=5$ (in the units
$d/\alpha$).} \label{fig2}
\end{figure}

\newpage
\begin{figure}
  \centering
  \includegraphics[width=165mm]{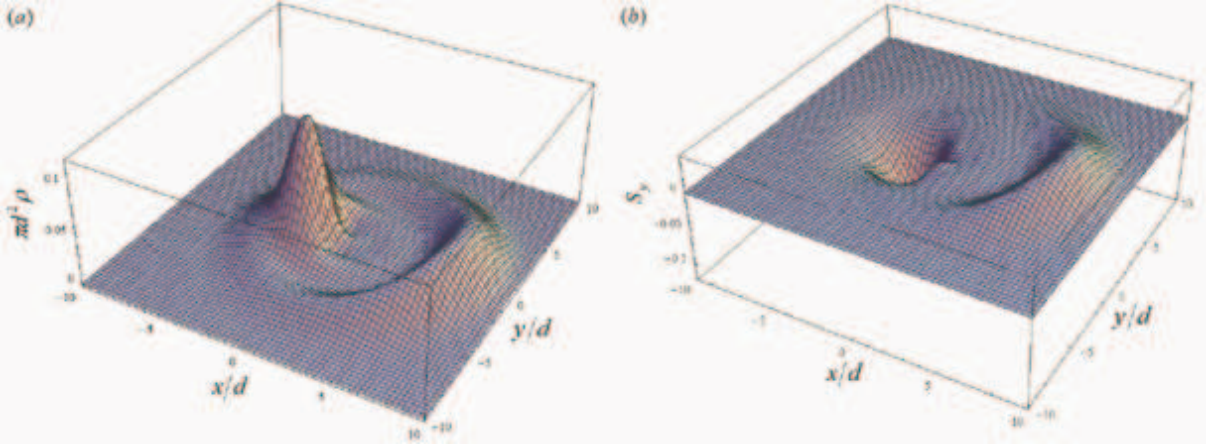}
\caption{(Color on line). Electron density
$\rho(x,t)=|\Psi_1|^2+|\Psi_2|^2$(a) and spin density $s_z(x,y,t)$
(b) for $k_0d=1$ at the moment $t=5$ (in units $d/\alpha$).}
\label{fig3}
\end{figure}

\newpage
\begin{figure}
  \centering
  \includegraphics[width=85mm]{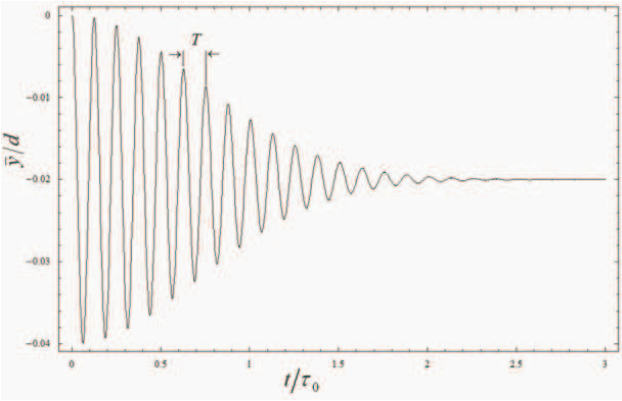}
\caption{The average coordinate of $\bar y(t)$ versus time for the
packet with $k_0d=25$.} \label{fig4}
\end{figure}

\newpage
\begin{figure}
  \centering
  \includegraphics[width=165mm]{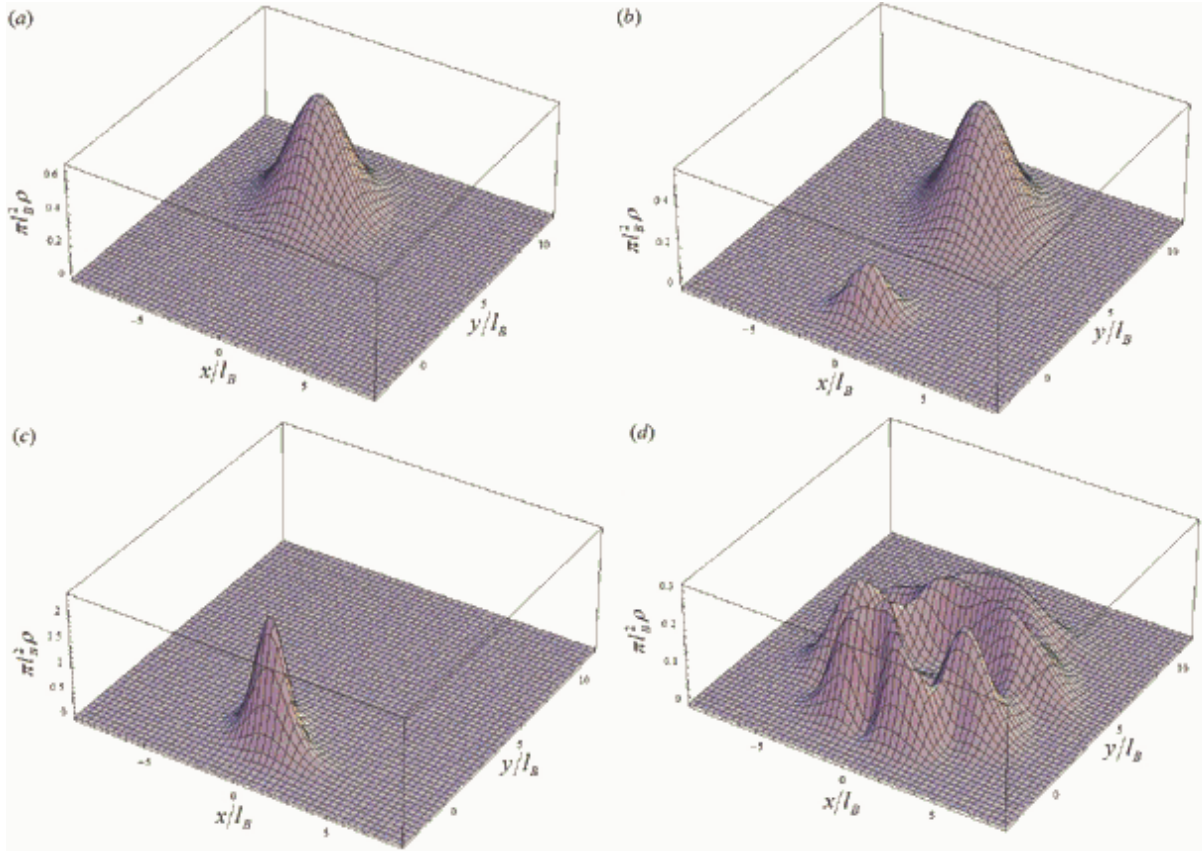}
\caption{(Color on line). Evolution of coherent wave packet
Eq.(31) in a perpendicular magnetic field: (a) the initial
electron density Eq.(31), (b) two split packets at time
$t_0\approx45\frac{2\pi}{\omega_c}$,(c) restored packets at time
$2t_0\approx90\frac{2\pi}{\omega_c}$, and (d) randomized electron
density for large time.} \label{fig5}
\end{figure}

\newpage
\begin{figure}
  \centering
  \includegraphics[width=85mm]{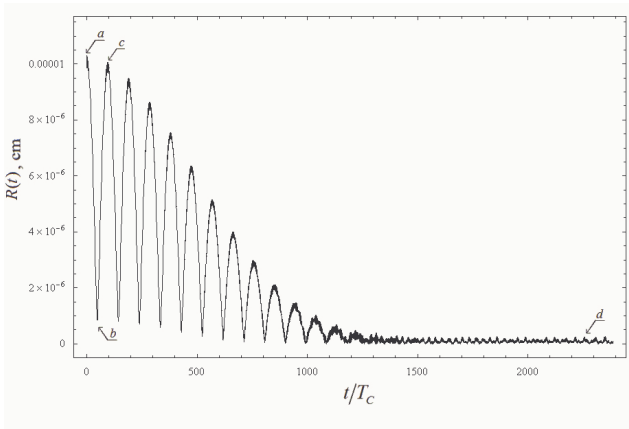}
\caption{Cyclotron radius plotted versus the time (for the same
parameters as in Fig.5). The distance between maximums and
minimums of $R(t)$ marked by arrows is approximately equal to
$80T_c$ Time is measured in units of cyclotron period
$T_c=\frac{2\pi}{\omega_c}$. The points {\it a}, {\it b}, {\it c},
{\it d} correspond to the same moments of time as in Fig.5(a),
(b), (c), (d), respectively.} \label{fig6}
\end{figure}

\end{document}